# Fast Bipartitioned Hybrid Adder Utilizing Carry Select and Carry Lookahead Logic

Padmanabhan Balasubramanian, Douglas L. Maskell
College of Computing and Data Science
Nanyang Technological University
50 Nanyang Avenue, Singapore 639798
(balasubramanian, asdouglas)@ntu.edu.sg

*Abstract*—**We present a novel fast bipartitioned hybrid adder (FBHA) that utilizes carry-select and carry-lookahead logic. The proposed FBHA is an accurate adder with a significant part and a less significant part joined together by a carry signal. In an N-bit FBHA, the K-bit less significant part is realized using carry-lookahead adder logic, and the (N – K)-bit significant part is realized using carry-select adder logic. The 32-bit addition was considered as an example operation for this work. Many 32-bit adders ranging from the slow ripple carry adder to the fast parallel-prefix Kogge-Stone adder and the proposed adder were synthesized using a 28-nm CMOS standard cell library and their design metrics were compared. A well-optimized FBHA achieved significant optimizations in design metrics compared to its high-speed adder counterparts and some examples are mentioned as follows: (a) 19.8% reduction in delay compared to a carry-lookahead adder; (b) 19.8% reduction in delay, 24.4% reduction in area, and 19.4% reduction in power compared to a carry-select adder; (c) 45.6% reduction in delay, and 13.5% reduction in power compared to a conditional sum adder; and (d) 46.5% reduction in area, and 29.3% reduction in power compared to the Kogge-Stone adder.**

## I. INTRODUCTION

High-speed adders offer many advantages and find wide-ranging applications. They execute arithmetic operations swiftly boosting the overall speed and efficiency of digital systems. High-speed adders minimize computation time, crucial for real-time digital signal processing applications. These adders enable computational systems to handle large volumes of data swiftly, enhancing overall throughput in applications such as data centres. With their rapid processing capabilities, high-speed adders help minimize delays in critical systems like communication networks. They play a crucial role in accelerating computations in high-performance computing environments, enabling faster simulations and scientific calculations. In applications like image and video processing, high-speed adders enable fast real-time processing of high-definition multimedia content, enhancing the user experience. High-speed adders are also employed in critical systems in automotive and aerospace applications, including sensor data processing, navigation, and control systems, ensuring rapid and precise operation.

Adders and multipliers play integral roles within the data paths of digital signal processing units. Within the spectrum of real-time digital signal processing benchmarks, addition stands out as the most commonly performed arithmetic operation [1]. Research conducted on the DLX prototype RISC machine [2] demonstrated that approximately 72% of its instructions involved addition or subtraction operations. Similarly, an analysis of operations handled by an ARM processor's arithmetic and logic unit indicated that additions represented nearly 80% of its computational workload [3]. Given these, the design of a high-speed adder with low power is important for digital circuits and systems.

Several adder architectures have been documented in the literature [4], including but not limited to the ripple carry adder, carry skip adder, conditional sum adder, carry-select adder, carry-lookahead adder, and a variety of parallel-prefix adders such as the Brent-Kung adder, Sklansky adder, Kogge-Stone adder, etc. In the subsequent discussion, we will provide a concise overview of each of these adder architectures.

The Ripple Carry Adder (RCA) is a basic one that is constructed through a series connection of one-bit full adders where the carry output of each full adder serves as the carry input for the successive full adder. The RCA is typically synthesized using a synthesis tool corresponding to an addition described in dataflow style using a hardware description language (HDL). In an RCA, addition is performed sequentially starting from the least significant input bit up to the most significant input bit. Therefore, the primary drawback of an RCA lies in the long propagation delay induced by the sequential carry propagation, constraining its operational speed. However, an RCA has advantages in terms of minimal area usage and power consumption. An alternative variant of the RCA, introduced in [5], substitutes the traditional cascade of one-bit full adders with a cascade of two-bit full adders. As noted in [5], this modification led to a significant improvement in the speed of an RCA compared to a conventional RCA (using one-bit full adders) albeit at the expense of increases in area and power.

The Carry Skip Adder (CSKA) [6], also called the carry bypass adder, tends to increase the addition speed by skipping carry propagation through specific bit groups. This is accomplished when carry propagation is constrained within specific bit groups and there is no carry propagation happening across different bit groups. Thus, the speed of a CSKA is data-dependent. Supposing, the inputs result in carry propagation across different bit groups then the best-case speed of a CSKA cannot be realized and the speed of a CSKA would resemble the speed of an RCA. Compared to an RCA, a CSKA however, requires extra logic to realize the carry-skip mechanism.

The Conditional Sum Adder (CSA) [7] optimizes addition by selectively generating carry signals based on input bits. This is achieved by incorporating some complex logic within a CSA. The CSA can predict carry propagation by analyzing

This research was partially funded by the Ministry of Education, Singapore Academic Research Fund under grant numbers Tier-1 RG48/21 and Tier-1 RG127/22.

the input bits and reducing unnecessary carry propagation. Thus, a CSA dynamically controls carry signals and can efficiently handle carry propagation when necessary thereby reducing the delay and improving the addition speed. In scenarios where carry propagation is sparse, CSAs can accelerate addition operations. The effectiveness of CSAs depends on the distribution of carry within input bits. CSAs typically tend to achieve optimal performance when the carry is concentrated within specific bit groups.

The Carry-Select Adder (CSLA) [8] is a high-speed adder that typically consists of groups of parallel adders each capable of generating its sum and carry output for a specific carry input condition. Among the parallel adders, one adder produces sum and carry output assuming a carry input of 0, and the other adder produces sum and carry output assuming a carry input of 1. The parallel adders of the CSLA are conventionally realized using the RCA architecture. However, an alternative realization of the CSLA is available whereby the sum and carry output produced by one adder for a carry input of 0 is increased by 1 through a corresponding add-one or binary to excess-1 code converter circuit [9]. The final sum and carry output of the CSLA are produced by selecting the correct sum and carry output among the outputs of parallel adders via 2-to-1 multiplexers whose select signal is the actual carry input. Given the use of parallel adders or the use of one adder and an add-one circuit to realize a CSLA, its logic complexity would be significant. However, despite this, CSLA offers a high-speed performance. Nevertheless, the hardware complexity and area overhead of a CSLA should be carefully managed for their deployment in real-world scenarios.

The Carry-Lookahead Adder (CLA) [10] is another high-speed adder that precomputes carry signals corresponding to different addition stages based on the input bits. In a CLA, the carry signals are determined independently of previous stages in parallel, and this helps to reduce the time required for carry propagation. CLAs have been implemented at the gate level and transistor level and they have been used in microprocessors and digital signal processors to accelerate addition operations. The carry-lookahead functionality is inherently present in field programmable gate arrays (FPGAs), which help to implement high-speed adders in an FPGA. CLAs generally offer superior performance compared to other high-speed adders such as CSKA, CSLA, etc. There are a couple of standard CLA implementations available namely, the conventional or regular CLA and the block CLA [11]. For an ASIC-based standard cell-based synthesis, Ref. [12] showed that the conventional CLA is preferable to a block CLA in terms of speed. Ling presented a variant of the regular CLA architecture, called the Ling adder [13].

The Parallel-Prefix Adder (PPA) [14] encompasses an advanced, tree-like architecture that consists of interconnected stages where partial addition and carry generation are performed simultaneously. Despite its complexity, PPA offers high-speed performance. PPAs utilize prefix computation techniques to effectively distribute the carry signals across the adder thus reducing the critical path delay. PPAs have been used in high-performance computing, microprocessors, and digital signal processors to accelerate arithmetic operations.

There are many types of PPAs such as the Brent-Kung adder (BKA) [15], Sklansky adder [16], Kogge-Stone adder (KSA) [17], etc. each having unique characteristics and offering trade-offs between design metrics such as delay and area/power. Hence, the implementation of PPAs requires careful consideration of hardware resources and design trade-offs while aiming to achieve optimal performance. For example, the BKA reduces the carry propagation via a balanced structure. The Sklansky adder incorporates a recursive tree structure enabling efficient carry computation. The Sklansky adder features a regular structure that is scalable and easily implementable but it suffers from a high delay for long adder chains due to serial carry propagation. The KSA has a dense tree structure and features excellent parallelism thus significantly reducing the delay. But the KSA's structure is irregular which complicates its layout. Another PPA called the Ladner-Fischer Adder (LFA) [18] strikes a balance between parallelism and regularity thus offering a compromise between complexity and performance. However, the LFA may not achieve the same level of optimization as other PPAs. The Han-Carlson Adder [19] utilizes a modified carry-skip technique to reduce delay thus improving the performance. However, its implementation complexity may limit its practicality in certain applications.

The remaining portion of the paper is organized as follows. Section II describes the proposed high-speed adder, and presents its variants. Section III discusses the synthesis method used to implement various adders including the proposed adder and presents their standard design metrics. Lastly, Section IV draws some conclusions from this research.

## II. Proposed Adder Architecture

We propose a novel Fast Bipartitioned Hybrid Adder called FBHA that involves carry-select and carry-lookahead logic, reflecting its hybrid nature. The architecture of FBHA is shown in Fig. 1. FBHA is a high-speed accurate adder that consists of two parts viz. a significant part that is realized using carry-select logic that is shown enclosed within the dashed brown box in Fig. 1, and a less significant part that is realized using carry-lookahead logic that is shown enclosed within the dashed blue box. In an N-bit FBHA, K bits are allotted to the less significant part, and (N – K) bits are allotted to the significant part. Thus, an N-bit FBHA comprises a K-bit CLA and an (N – K)-bit CSLA.

The K-bit CLA is realized via a cascade of small-size CLA modules which may be of the same or different sizes. For example, considering a 32-bit FBHA, and assuming K = 24, a 24-bit CLA may be realized by a cascade of six 4-bit CLA modules (or) one 8-bit CLA module, three 4-bit CLA modules, and two 2-bit CLA modules. Given this realization, the RCAs comprising the 8-bit CSLA would consist of 7 full adders and 1 half adder. The choice of CLA modules would tend to influence the CLA's delay and eventually the FBHA's delay. Therefore, it is important to choose CLA modules such that the resulting FBHA would be delay-optimized. The carry output generated by the CLA is forwarded to the CSLA and is used as the select signal for the multiplexers present in the CSLA.

The CSLA consists of two RCAs one with 0 as the carry input and another having a carry input of 1. The RCAs present in the CSLA can add the most significant input bits in parallel. Each of the sum bits produced by the two RCAs is given to a 2-to-1 multiplexer (Mux21), whose select signal is the carry output produced by the CLA. Depending on the value of the select signal, the correct sum of the CSLA is selected. The sum output by the CSLA and CLA when concatenated represents the final sum of the FBHA.

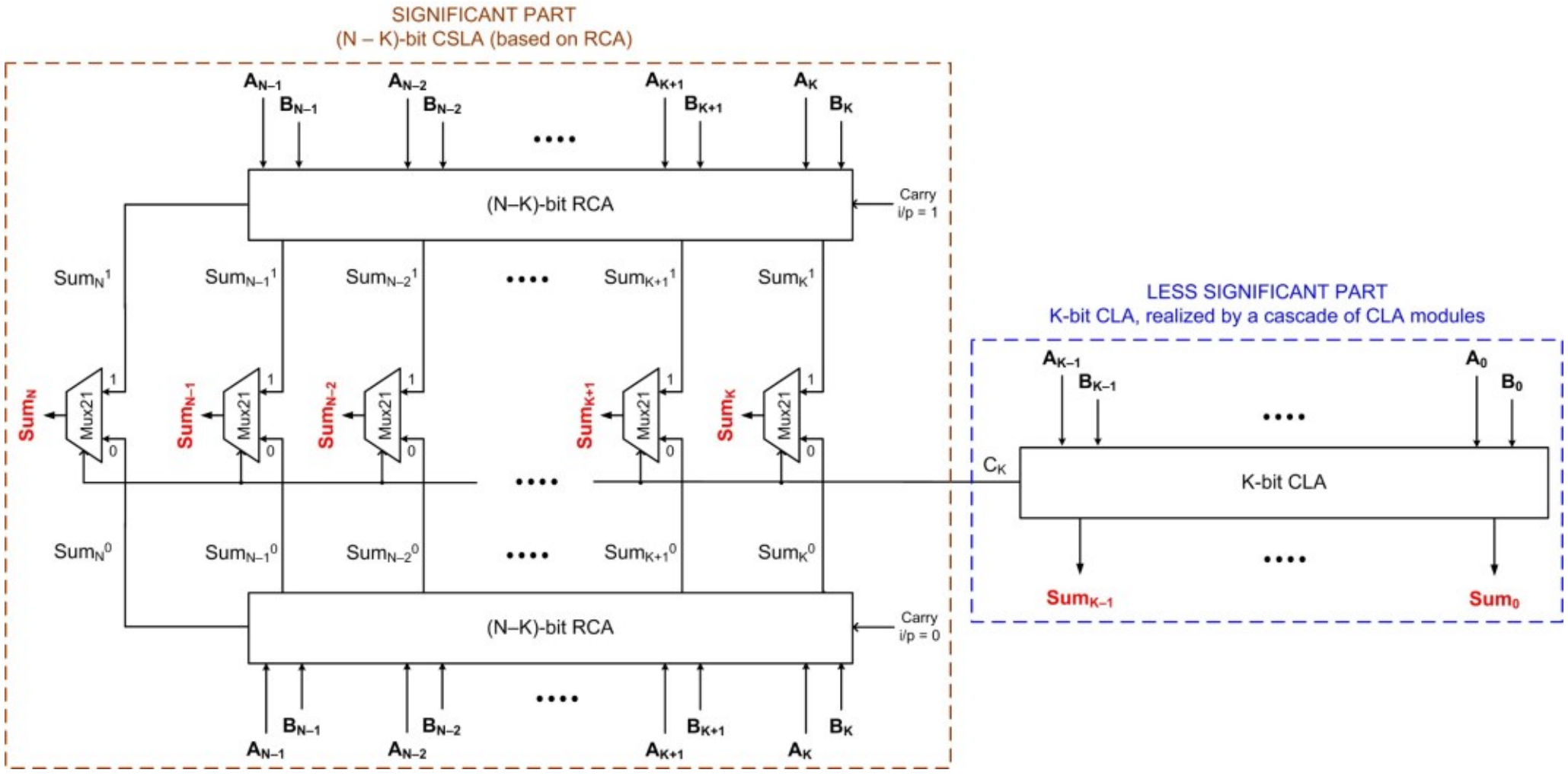

Fig. 1. Architecture of proposed N-bit fast bipartitioned hybrid adder (FBHA). A and B represent the adder inputs; Sum represents the adder output.

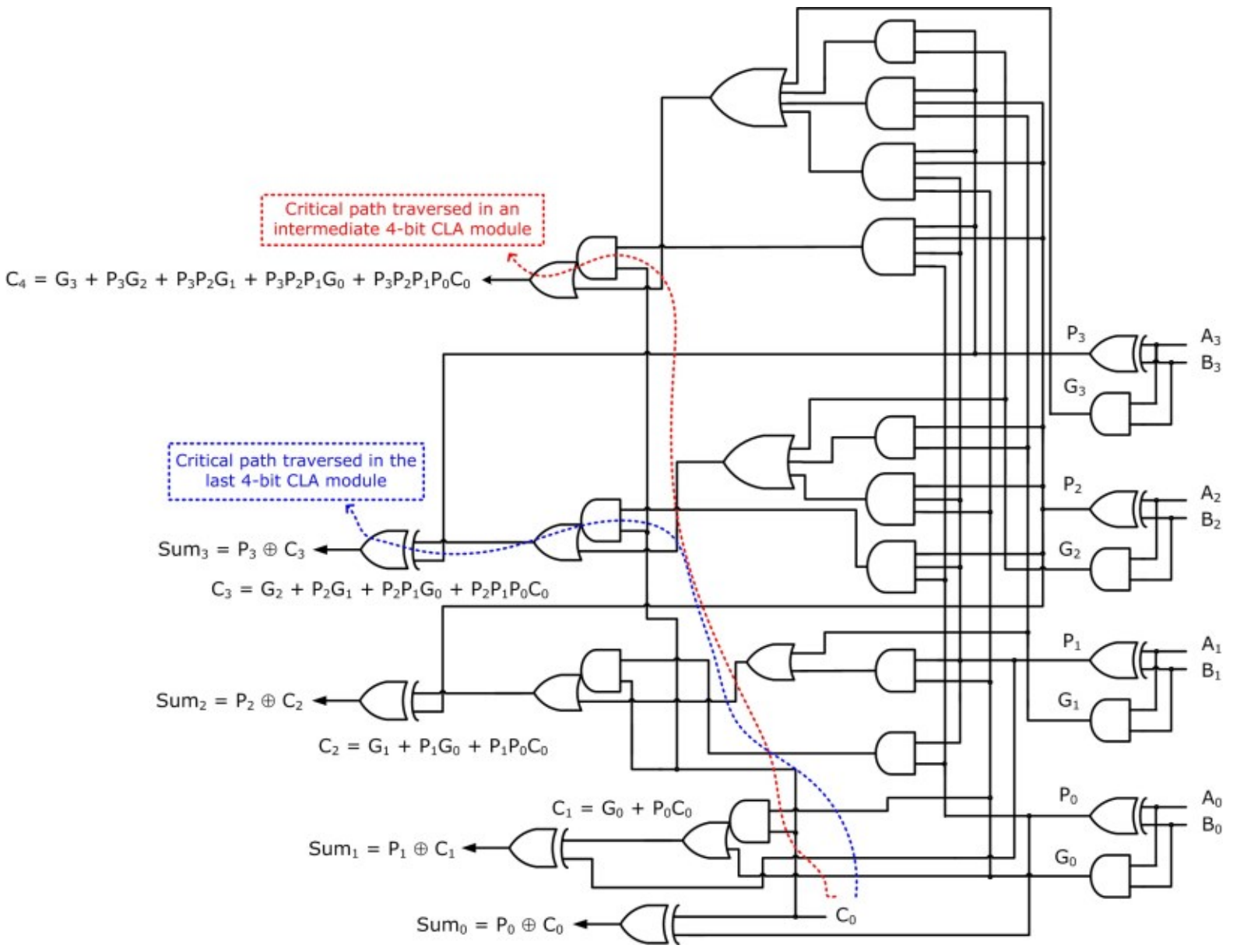

Fig. 2. Gate-level representation of a 4-bit delay-optimized CLA module with carry input.

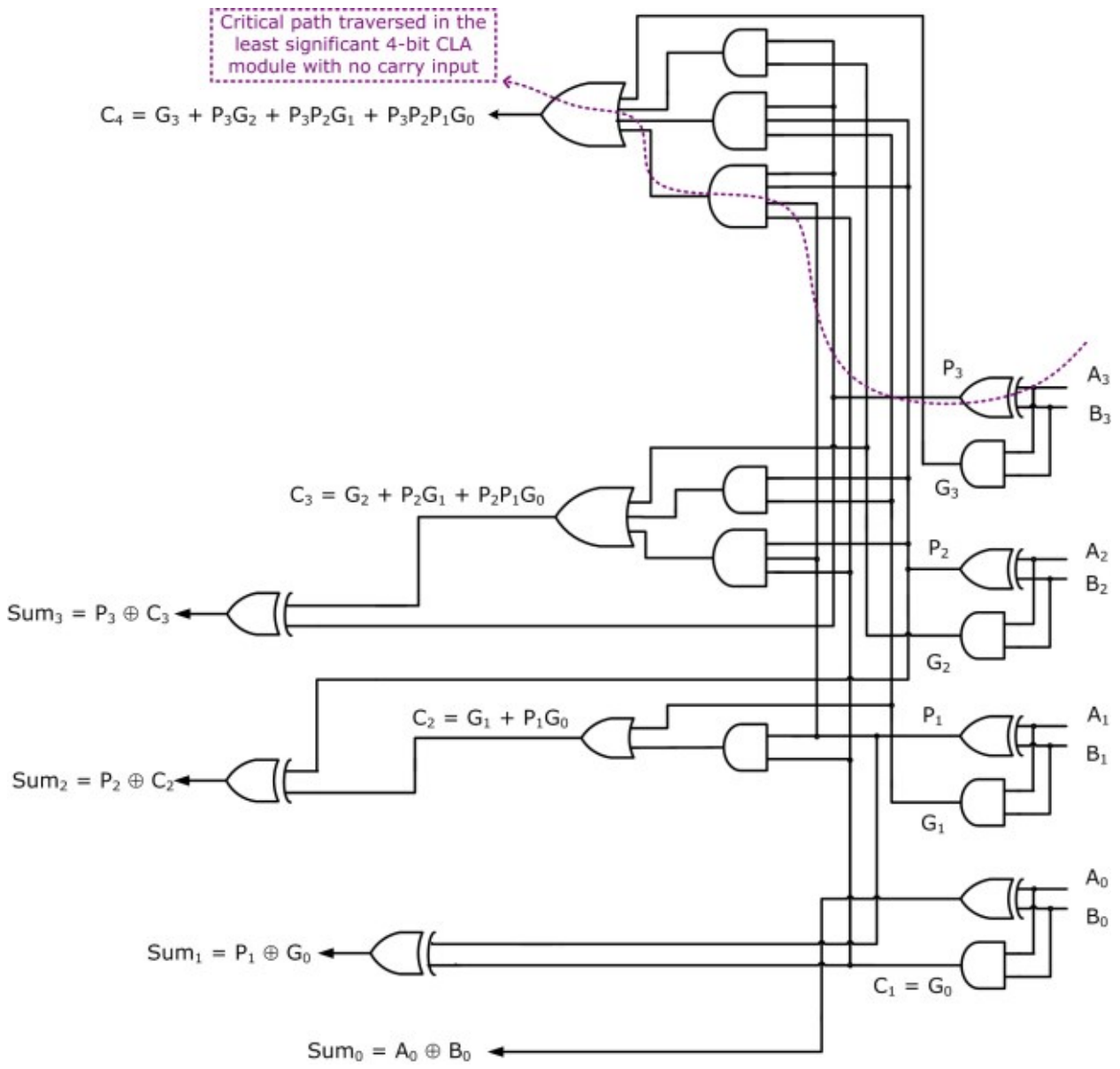

Fig. 3. Gate-level representation of a 4-bit CLA module with no carry input.

In general, compared to an RCA which has a linear delay, a CLA is preferable as it has a logarithmic delay. Therefore, compared to the RCAs comprising the CSLA, the CLA is faster. Hence, it is important to ensure that the speed of the CLA dictates the FBHA speed. This implies the CLA size should be greater than the CSLA size so that the FBHA delay would be defined by the delay of the K-bit CLA plus a Mux21 delay that corresponds to the (N – K)-bit CSLA. As a result, an N-bit FBHA is likely to be faster compared to an N-bit CLA or CSLA.

The K-bit CLA comprising the N-bit FBHA in Fig. 1 is better realized by a cascade of delay-optimized CLA modules. Ref. [20] presented a delay-optimized CLA module realization that is faster than a conventional CLA module realization, and hence we utilized that for this work. For example, the gate-level realization of a delay-optimized 4-bit CLA module (with a carry input) is shown in Fig. 2. In the absence of a carry input, the logic of the 4-bit CLA module would be reduced, as shown in Fig. 3. In Figs. 2 and 3, ($A_3$, $A_2$, $A_1$, $A_0$) and ($B_3$, $B_2$, $B_1$, $B_0$) denote the adder inputs, $C_0$ represents the carry input, and ($C_4$, $C_3$, $C_2$, $C_1$) represent the lookahead carry outputs. ($Sum_3$, $Sum_2$, $Sum_1$, $Sum_0$) denote the sum output by the 4-bit CLA. The equations for sum and carry outputs of the 4-bit CLA module with/without the carry input are also given in Figs. 2 and 3. Let us assume that the 4-bit CLA modules shown in Figs. 2 and 3 are used to realize a K-bit CLA forming a part of an N-bit FBHA and that K modulo 4 = 0, and (K/4) > 3. Given this scenario, the critical path that will be traversed in the first CLA module is highlighted by the dotted violet line in Fig. 3. The critical path that will be traversed in an intermediate CLA module is highlighted by the dotted red line in Fig. 2, and the critical path that will be traversed in the last CLA module is highlighted by the dotted blue line in Fig. 2.

## III. Synthesis and Results

We considered the 32-bit addition as a case study for this work. Given this, first, it is important to determine the optimum size of CLA and CSLA that form a part of FBHA. Toward this, we structurally described 32-bit FBHAs containing different sizes of CSLA and CLA in Verilog HDL as follows: (i) a 4-bit CSLA and a 28-bit CLA, (ii) an 8-bit CSLA and a 24-bit CLA, (iii) a 12-bit CSLA and a 20-bit CLA, and (iv) a 16-bit CSLA and a 16-bit CLA. The CLAs comprising the different FBHAs were realized by a cascade of delay-optimized 4-bit CLA modules, shown in Fig. 2.

To synthesize the FBHAs using gates, we used Synopsys DesignCompiler and a 28-nm CMOS standard cell library [21]. A typical case PVT specification of the library was considered, having a supply voltage of 1.05 V and an operating junction temperature of 25 °C. During synthesis, the default wire-load model was invoked and a fanout-of-4 drive strength was assigned to all the output ports i.e., the sum bits. Following synthesis, the gate-level netlists were subjected to functional simulation using Synopsys VCS. To do this, a test bench containing approximately one thousand random input vectors was uniformly applied to all adders at a latency of 4 ns. The switching activity data obtained from functional simulations was used to estimate the total (average) power dissipation using PrimePower. PrimeTime was used to assess the critical path delay, and DesignCompiler was used to estimate the overall area of the adders, including cell and interconnect areas.

The design metrics of 32-bit FBHAs containing different sizes of CSLA and CLA are given in Table I. In Table I, the notation 'FBHA_X_Y' refers to a 32-bit FBHA that consists of an X-bit CSLA and a Y-bit CLA, and X + Y = 32.

TABLE I. Design metrics of 32-bit FBHAs containing different sizes of CSLA and CLA

| FBHA and Versions | Area ($\mu m^2$) | Delay (ns) | Power (µW) |
|---|---|---|---|
| FBHA_4_28 | 566.13 | 1.09 | 54.64 |
| FBHA_8_24 | 623.39 | 1.03 | 59.52 |
| FBHA_12_20 | 677.25 | 1.14 | 63.62 |
| FBHA_16_16 | 729.96 | 1.41 | 68.24 |

As mentioned previously, the CSLA and CLA sizes should be chosen such that the FBHA delay should be optimum and primarily dictated by the CLA delay i.e., the FBHA delay should be equal to the sum of the CLA delay and a Mux21 delay (corresponding to the CSLA). Given this, it is observed from Table I that when the CLA size is reduced from 28 bits to 24 bits, the FBHA delay is also reduced. This is because the CSLA delay is subsumed in the CLA delay, and since the number of 4-bit CLA stages is reduced from 7 in the case of FBHA_4_28 to 6 in the case of FBHA_8_24, thus the delay of FBHA_8_24 is less compared to the delay of FBHA_4_28. However, when the CLA size is reduced further from 24 bits to 20 bits, the FBHA delay increases. This is because the CSLA delay tends to dominate the CLA delay. When the CLA size is reduced further from 20 bits to 16 bits, the FBHA delay further increases, which is expected due to the dominance of the CSLA delay i.e., by the delay of the RCA in the CSLA. Thus, any further reduction in the CLA size would negatively impact the FBHA delay, and so further reductions in CLA sizes were not considered. Among the four FBHAs shown in Table 1, FBHA_8_24 is found to be optimum from the delay perspective.

FBHA_8_24 can be realized in different ways i.e., the 24-bit CLA portion comprising the FBHA can be implemented using same-size or different-size CLA modules. Table II shows the design metrics of different versions of FBHA_8_24 with the 24-bit CLA constituent realized via a cascade of same-size or different-size CLA modules as follows: (i) twelve 2-bit CLA modules (denoted by FBHA_2×12), (ii) six 4-bit CLA modules (denoted by FBHA_4×6), (iii) four 6-bit CLA modules (denoted by FBHA_6×4), (iv) three 8-bit CLA modules (denoted by FBHA_8×3), (v) one 8-bit CLA module and four 4-bit CLA modules (denoted by FBHA_84444), (vi) two 6-bit CLA modules and three 4-bit CLA modules (denoted by FBHA_66444), (vii) one 8-bit CLA module, three 4-bit CLA modules, and two 2-bit CLA modules (denoted by FBHA_844422), and (viii) two 6-bit CLA modules, two 4-bit CLA modules, and two 2-bit CLA modules (denoted by FBHA_664422). Although we synthesized many 32-bit FBHAs comprising a 24-bit CLA using different-size CLA modules, we have shown only four variants in Table II which were found to be better optimized in terms of delay.

Table II shows that among the FBHAs using same-size CLA modules FBHA_4×6 and FBHA_6×4 have the same delay and are close in terms of area and power dissipation. But, we also see from Table II that FBHAs comprising different-size CLA modules are better optimized in terms of delay compared to FBHAs comprising same-size CLA modules. Therefore, the former is preferable. Among the FBHAs comprising different-size CLA modules,

FBHA_844422 and FBHA_664422 are slightly better compared to FBHA_84444 and FBHA_66444 in terms of delay, and so they are preferable. Moreover, FBHA_844422 and FBHA_664422 have almost the same area and power. In terms of the power-delay product (PDP) that serves as a representative figure of merit for energy efficiency, FBHA_844422 is found to be better optimized compared to its counterparts given in Table II.

TABLE II. DESIGN METRICS OF DIFFERENT FBHA_8_24 WITH THE CLA PORTION REALIZED USING SAME- OR DIFFERENT-SIZE CLA MODULES

| FBHA_8_24 Variants | Area (μm$^2$) | Delay (ns) | Power (μW) |
|---|---|---|---|
| FBHA_2×12 | 617.89 | 1.36 | 59.41 |
| FBHA_4×6 | 623.39 | 1.03 | 59.52 |
| FBHA_6×4 | 619.26 | 1.03 | 59.33 |
| FBHA_8×3 | 612.22 | 1.10 | 58.69 |
| FBHA_84444 | 626.69 | 0.96 | 59.68 |
| FBHA_66444 | 626.56 | 0.96 | 59.87 |
| FBHA_844422 | 635.99 | 0.93 | 60.07 |
| FBHA_664422 | 635.87 | 0.93 | 60.26 |

To make a comparison between our proposed adder and existing adders belonging to different architectures based on the design metrics, we implemented many 32-bit adders by following the same synthesis methodology described earlier. Besides the 32-bit FBHAs mentioned above, 32-bit versions of CSKA, CSLA, CLA, and some PPAs were structurally described in Verilog HDL (at the gate level) and synthesized using the same standard cell library for the same typical case PVT specification. The 32-bit RCA was synthesized based on a 32-bit addition described in dataflow style in Verilog using the addition operator (+). The 'compile_ultra' command was used to synthesize the RCA and the 'compile' command with speed specified as the optimization goal was used to synthesize the high-speed adders using DesignCompiler. The 32-bit CSKA was implemented via a cascade of eight 4-bit CSKA blocks. Concerning the CLA, two types were implemented – one being the conventional CLA (referred to as CCLA here) and another being the delay-optimized CLA (referred to as DCLA here). CCLA is composed of conventional CLA modules and DCLA is composed of delay-optimized CLA modules. The difference between a conventional 4-bit CLA module and a delay-optimized 4-bit CLA module is illustrated in [20] – an interested reader may refer to the same for details. In [20], a 32-bit CCLA and a 32-bit DCLA were implemented using respective 4-bit CLA modules, and we adopted the same for this work.

The basic Synopsys DesignWare library contains synthesizable RTL models of high-speed adders like the Ling adder (which is a CLA variant), CSA, and a couple of PPAs viz. the BKA and the Sklansky adder. We invoked all these high-speed adders for synthesis using DesignCompiler. Further, we used the structural description of a 32-bit KSA presented in [22] for synthesis. Concerning the CSLA, [20] analyzed four popular configurations, and found that the 32-bit CSLA featuring a uniform 8-8-8-8 input partition was better optimized for delay. Therefore, we considered this CSLA type alone for this work. The design metrics of different 32-bit adders, including the proposed and better-optimized FBHA are given in Table III.

TABLE III. DESIGN PARAMETERS OF VARIOUS 32-BIT ADDERS, SYNTHESIZED USING A 28-NM CMOS STANDARD DIGITAL CELL LIBRARY

| Adder | Area (μm$^2$) | Delay (ns) | Power (μW) |
|---|---|---|---|
| RCA | 166.01 | 3.40 | 42.13 |
| CSKA | 452.39 | 2.93 | 44.23 |
| CSA | 490.13 | 1.71 | 69.43 |
| CSLA | 841.01 | 1.16 | 79.45 |
| CCLA | 390.36 | 2.56 | 42.00 |
| Ling adder | 467.61 | 2.39 | 67.48 |
| DCLA | 513.82 | 1.16 | 49.29 |
| BKA | 484.25 | 2.42 | 56.65 |
| Sklansky adder | 449.81 | 2.74 | 57.10 |
| KSA | 1188.72 | 0.73 | 84.99 |
| FBHA_844422 | 635.99 | 0.93 | 60.07 |

Some important observations can be made from Table III. In terms of area, RCA is the best-optimized and this is expected because the RCA comprises just 31 full adders and 1 half adder. Consequently, RCA reports less power dissipation compared to many other adders, and only the CCLA has a comparable power dissipation. However, the RCA reports a 2.7× increase in delay compared to FBHA_844422. In terms of delay, excepting the KSA, the proposed FBHA_844422 reports reduced delay compared to the RCA, CSKA, CSA, CSLA, CCLA, Ling adder, DCLA, BKA, and the Sklansky adder. Nonetheless, there are only two 32-bit adders which are sub-ns adders i.e., adders having less than 1 ns delay which are KSA and FBHA_844422. Compared to FBHA_844422, KSA has a 21.5% reduced delay. However, FBHA_844422 requires 46.5% less area and dissipates 29.3% less power compared to the KSA.

The PDP of all the adders shown in Table III was calculated and normalized, and the normalized PDP values are plotted in Fig. 4. To do the normalization, the actual PDP of each adder was divided by the highest PDP among the lot, which corresponds to the Ling adder. Generally, it is desirable to minimize power and delay, and hence it is desirable to minimize the (normalized) PDP as well. Given this, a normalized PDP value of 1 represents an inferior design while the adder having the least normalized PDP value indicates a preferred design.

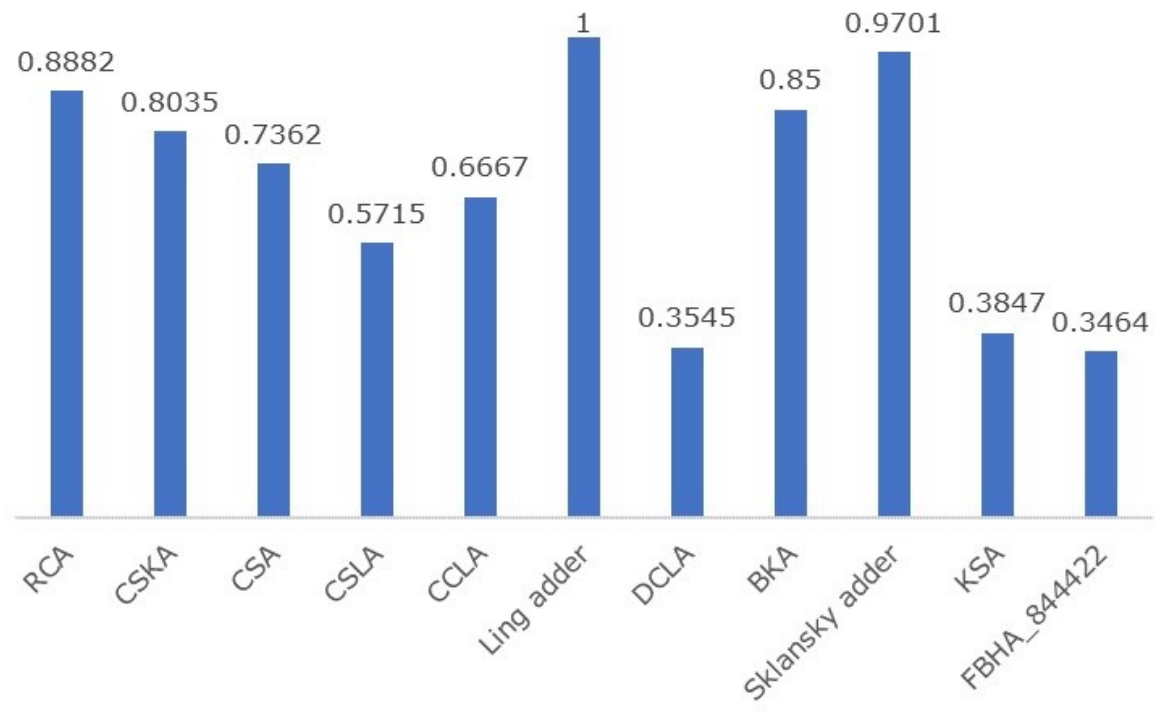

Fig. 4. Normalized power-delay product (PDP) of various 32-bit adders; the least value corresponds to the proposed adder viz. FBHA_844422.

From Fig. 4, it is seen that the PDP of DCLA and the proposed FBHA_844422 are less and comparable although the latter reports a 2.3% reduction. Nonetheless, FBHA_844422 has a 19.8% reduced delay compared to DCLA. With KSA and the proposed adder being sub-ns adders, FBHA_844422 features a 10% reduced PDP than KSA. The area generally merits less consideration than delay and power in modern VLSI designs. Thus, from the perspectives of delay and power, FBHA_844422 is considered preferable to its counterparts.

## IV. Conclusion

This paper presented a new accurate fast bipartitioned hybrid adder (FBHA). The proposed adder has two parts namely, a significant part and a less significant part. With N bits representing the FBHA size, and K bits representing the size of the less significant part, the size of the significant part would be (N – K) bits. The hybrid nature of the proposed adder stems from the fact that its significant and less significant parts are realized using two different adder architectures i.e., the significant part is realized using the CSLA architecture, and the less significant part is realized using the CLA architecture. The interesting aspect of the FBHA is that the CSLA delay is meant to be subsumed in the CLA delay, and the sizes of CLA and CSLA should be chosen accordingly. Since the CLA size is less than N bits, therefore, the FBHA delay would be potentially less than the delay of an N-bit CLA or CSLA.

To compare the performance of the proposed adder with other existing adders, the 32-bit addition was considered as an example operation. Many adders were implemented following a semi-custom ASIC design style using a 28-nm CMOS digital cell library. The CSLA unit of the FBHA was implemented using RCAs, and the CLA unit of the FBHA was implemented via a cascade of same-size or different-size CLA modules. The synthesis results show that it is best to implement the CLA unit of the FBHA using a cascade of different-size CLA modules. Among the many FBHAs realized, one variant namely FBHA_844422 was found to be better optimized for delay. The synthesis results showed that the KSA and FBHA_844422 are sub-ns adders. Nevertheless, FBHA_844422 requires less area, dissipates less power, and has a lesser PDP than the KSA. Moreover, FBHA_844422 is reportedly faster than many of its counterparts. Therefore, from the combined perspectives of delay, power, and PDP, the proposed FBHA (FBHA_844422) is considered preferable. As further work, the efficacy of FBHA in realizing other arithmetic operations such as multiplication in a high-speed and energy-efficient manner may be investigated. Also, the utility of the FBHA to efficiently realize higher-width additions may be studied as future work.